\documentclass{iaus}
\usepackage{graphicx}

\def\mc{meridional circulation}
\def\cz{convection zone}

\title[Dynamo models of grand minima] 
{Dynamo models of grand minima}

\author[Choudhuri]   
{Arnab Rai Choudhuri}

\affiliation{Department of Physics, Indian Institute of Science, Bangalore-560012\\ email: {\tt arnab@physics.iisc.ernet.in}} 

\pubyear{2010}
\volume{xxx}  
\pagerange{1--6}
\jname{Title of your IAU Symposium}
\editors{A.C. Editor, B.D. Editor \& C.E. Editor, eds.}
\begin{document}

\maketitle

\begin{abstract}

Since a universally accepted dynamo model of grand minima does not exist at the
present time, we concentrate on the physical processes which may be behind the
grand minima. After summarizing the relevant observational data, we make the
point that, while the usual sources of irregularities of solar cycles may be
sufficient to cause a grand minimum, the solar dynamo has to operate somewhat
differently from the normal to bring the Sun out of the grand minimum. We then
consider three possible sources of irregularities in the solar dynamo: (i) nonlinear
effects; (ii) fluctuations in the poloidal field generation process; (iii) fluctuations in
the meridional circulation.  We conclude that (i) is unlikely to be the cause
behind grand minima, but a combination of (ii) and (iii) may cause them. If
fluctuations make the poloidal field fall much below the average or make the
meridional circulation significantly weaker, then the Sun may be pushed into
a grand minimum.
 
\keywords{Sun: dynamo --- Sun: activity --- sunspots}
\end{abstract}

\firstsection 


\section{Introduction}

At the very outset, I would like to mention that the subject of this invited talk
was not chosen by me.  The organizers felt that this symposium should have an
invited talk on dynamo models of grand minima and requested me to give it.
Only after considerable initial hesitation, I finally agreed.
The reason behind my initial hesitation is that at present we have no dynamo
model of grand minima which is completely satisfactory or which is generally
accepted in the community.  No two
self-respecting dynamo theorists seem to agree how grand minima are produced!
To the best of my knowledge, this is the first time an invited talk on this subject
is being given in a major international conference.

Given this situation, I have decided to adopt the following strategy. I shall
mainly focus on the various bits of physics which go into making models of grand
minima rather than discussing specific models of grand minima in detail. While
many of the present-day models of grand minima may eventually fall by the wayside,
I believe that the bits of physics that we consider relevant today will still
remain relevant after 20 or 30 years when there may be a better understanding
of what occasionally pushes the Sun into the grand minima. 

\section{Observational characteristics of grand minima}

Before getting into the theoretical discussion, let us see what we can learn
about the characteristics of grand minima from the very limited observational
data available to us.

Several authors have studied the archival records of sunspots during the Maunder minimum
(Sokoloff \& Nesme-Ribes 1994; Hoyt \& Schatten 1996). The few sunspots seen during the Maunder
minimum mostly appeared in the southern hemisphere. Sokoloff \& Nesme-Ribes (1994) have used the
archival data to construct a butterfly diagram for a part of the Maunder minimum from 1670,
showing a clear trend of hemispheric asymmetry. It is an open question whether hemispheric
asymmetry played any crucial role in creating the Maunder minimum (Charbonneau 2005). Usoskin,
Mursula \& Kovaltsov (2000) argued that the Maunder minimum started abruptly but ended in a gradual
manner, indicating that the strength of the dynamo must be building up as the Sun came out of the
Maunder minimum. However, some recent evidence suggests that the onset of the Maunder minimum
may not be as abrupt as believed earlier (Vaquero et al.\ 2011).

When solar activity is stronger, magnetic fields in the solar wind suppress the cosmic
ray flux, reducing the production of $^{10}$Be and $^{14}$C which can be used as proxies for solar activity.
From the analysis of $^{10}$Be abundance in a polar ice core, Beer, Tobias \& Weiss (1998) concluded that
the solar activity cycle continued during the Maunder minimum, although the overall level of the activity
was lower than usual. Miyahara et al. (2004) drew the same conclusion from their analysis of $^{14}$C
abundance in tree rings.

This method of using various proxies (like the abundances of $^{10}$Be and $^{14}$C) 
for sunspot activity can be extended
to study even the earlier grand minima before the Maunder minimum.  Usoskin, Solanki \& Kovaltsov (2007)
estimated that there have been about 27 grand minima in the last 11,000 years. They also identified
about 19 grand maxima, i.e. periods during which sunspot activity was unusually high, like what was
seen during much of the twentieth century.

\section{Are grand minima merely extremes of cycle irregularities?}

We know that the solar cycle is only approximately periodic. Both the strength
and the period vary from one cycle to another. We begin our theoretical discussion
by raising the question whether the grand minima are merely extreme examples of
cycle irregularities.  Are the theoretical ideas used to model irregularities
of solar cycles adequate to explain the occurrences of grand minima, or do we
need to invoke some qualitatively different ideas?  We do not yet have a definitive
answer to this question.  Any dynamo theorist is entitled to have his or her own personal
opinion.  Let me put forth my personal opinion.

Our simulations (to be discussed later) seem to suggest that nothing very extraordinary
may be needed to push the Sun into a grand minimum.  If the fluctuations which
cause the usual cycle irregularities are sufficiently large, they may sometimes
cause grand minima.  However, even after the Sun is pushed into the grand minimum, 
a subdued cycle has to continue (as discussed in \S~2) and eventually the Sun
has to come out of the grand minimum.  These are more problematic to explain.
There are certain mechanisms of magnetic field generation which crucially depend
on the existence of sunspots.  Certainly those mechanisms cannot be operative
during a grand minimum.  So we need to invoke alternative mechanisms.

\def\vb{\bf v}
Let us look at the question how magnetic fields are generated in the dynamo process.
The basic idea of solar dynamo is that the toroidal and poloidal components of
the solar magnetic field sustain each other through a feedback loop.  It is
fairly easy to generate the toroidal field by the stretching of the poloidal field
due to differential rotation.  Since helioseismology has shown that the differential
rotation is concentrated in the tachocline, the generation of toroidal field 
mainly takes place there.  To complete the loop, we need to generate the poloidal
field from the toroidal field.  The historically important idea of Parker (1955)
--- which was further elaborated by Steenbeck, Krause \& R\"adler (1966) --- is
that the cyclonic turbulence in the convection zone twists the toroidal field to
produce the poloidal field.  This mechanism is often called the $\alpha$-effect
because the crucial parameter describing this process is usually denoted by the
symbol $\alpha$. Within certain approximation schemes, this parameter can be shown
to be given by
$$\alpha = - \, \frac{1}{3}\, \overline{\vb. (\nabla \times \vb)}\; \tau, \eqno(1)$$
where $\vb$ is the fluctuating part of the velocity field and $\tau$ is the correlation
time (see, for example, Choudhuri 1998, \S~16.5).  This $\alpha$-effect mechanism can be operative
only if the toroidal field is not too strong such that the helical turbulence is
able to twist it.  However, the flux tube rise simulations by several authors
(Choudhuri \& Gilman 1987; Choudhuri 1989; D'Silva \& Choudhuri 1993; Fan, Fisher
\& DeLuca 1993; Caligari et al.\ 1995) indicated that the toroidal field at the
base of the solar convection zone has to be as strong as $10^5$ G.  Such a strong
field cannot be twisted by helical turbulence and we cannot invoke $\alpha$-effect
to generate the poloidal field from such a strong toroidal field. An alternative
mechanism which has been widely used in many recent dynamo simulations is due to
Babcock (1961) and Leighton (1969). Bipolar sunspots on the solar surface have a
tilt with respect to the solar equator and this tilt increases with latitude.  This
was discovered by Joy in 1919 and is known as {\em Joy's law}. D'Silva \& Choudhuri
(1993) provided the first theoretical explanation of Joy's law by showing that the
tilt is produced by the Coriolis force acting on the flux tubes rising through the
convection zone due to magnetic buoyancy.  When a tilted bipolar sunspot decays, fluxes
of opposite polarities diffuse at slightly different latitudes, contributing to
the poloidal field.  According to this Babcock--Leighton mechanism, a tilted bipolar
sunspot pair is a conduit for converting the toroidal field to the poloidal field.
The sunspot pair forms due to the buoyant rise of the toroidal field and we get the poloidal
field after its decay.

Since we see the Babcock--Leighton mechanism clearly operational at the solar 
surface, most of the recent flux transport dynamo models take this as the primary
generation mechanism of the poloidal field.  The $\alpha$-effect cannot
operate on the strong toroidal field at the base of the convection zone. But
this strong toroidal field is expected to be highly intermittent (Choudhuri 2003)
and the $\alpha$-effect is likely to be
operative in those regions of the convection zone where the toroidal
field is weak, although the nature, the spatial distribution and even the algebraic
sign of the $\alpha$ parameter remain unclear at the present time.  The
Babcock--Leighton mechanism presumably cannot work during the grand minimum
when there are no sunspots.  So we have to fall back upon the $\alpha$-effect to
continue the cycles during the grand minimum and eventually to pull the Sun out
of it.  Our lack of knowledge about the $\alpha$-process limits our understanding
of these phenomena.  In the mean field dynamo equations, the term capturing the
Babcock--Leighton process is formally very similar to the term capturing the
$\alpha$-effect --- often even using the symbol $\alpha$.  Hence many dynamo models
of the grand minima are worked out at the present time by solving the same
equations during and outside the grand minima.  But it should be kept in
mind that the physics behind the symbol $\alpha$ must be very different during
and outside the grand minima.

In summary, our view is that the usual sources of irregularities in solar cycles are
sufficient for the onset of a grand minimum, but to pull the Sun out of a grand minimum
we need some physics different from the physics behind the usual solar cycles. Presumably
the situation is somewhat different for grand maxima.  Not only are the usual irregularities
expected to cause a grand maximum, we also do not require anything unusual to take the
Sun out of the grand maximum.  The usual poloidal field generation by the Babcock--Leighton
mechanism continues during the grand maximum.  It is intriguing that Usoskin, Solanki
\& Kovaltsov (2007) concluded that the lengths of grand maxima correspond to an exponential
distribution, but the lengths of grand minima have a more complicated bimodal distribution.
Is this connected with the fact that grand maxima do not involve any physical processes different
from the normal, but grand minima require the generation process of the poloidal field
to be different from the normal situation? 

\section{The origin of irregularities in the flux transport dynamo}    

We now discuss the possible sources of irregularities in the flux transport dynamo --- the
most widely studied model of the solar cycle in recent years. 
Let us begin by recapitulating some basic facts about the flux transport dynamo.
The toroidal field is generated in the tachocline by the strong differential rotation
and then rises to the surface due to magnetic buoyancy to form tilted bipolar sunspots.
When these sunspots decay, we get the poloidal field by the Babcock--Leighton mechanism.
The meridional circulation of the Sun, which is found to be poleward in the upper
layers of the convection zone and must have a hitherto unobserved equatorward branch
at the bottom of the convection zone in order to conserve mass, plays a very crucial
role in the flux transport dynamo. The \mc\ causes the observed poleward transport
of the poloidal field.  At the base of the \cz, it is responsible for making the
dynamo wave propagate equatorward, such that sunspots are produced at lower and
lower latitudes with the progress of a cycle.  In the absence of the \mc, the
dynamo wave at the bottom of the \cz\ would propagate poleward in accordance with
the Parker--Yoshimura sign rule (Parker 1955; Yoshimura 1975) contradicting observations.
The flux transport dynamo could become a serious model of the solar cycle only after
Choudhuri, Sch\"ussler \& Dikpati (1995) demonstrated that a sufficiently strong
\mc\ could overrule the Parker--Yoshimura sign rule and make the dynamo wave propagate
in the correct direction.

The original flux transport dynamo model of Choudhuri, Sch\"ussler
\& Dikpati (1995) led to two offsprings: a high diffusivity model
and a low diffusivity model.  The diffusion times in these two models
are of the order of 5 years and 200 years respectively.  The high
diffusivity model has been developed by a group working in IISc Bangalore
(Choudhuri, Nandy, Chatterjee, Jiang, Karak), whereas the low diffusivity
model has been developed by a group working in HAO Boulder (Dikpati,
Charbonneau, Gilman, de Toma).  The differences between these models
have been systematically studied by Jiang, Chatterjee \& Choudhuri
(2007) and Yeates, Nandy \& Mckay (2008).  Both these models are
capable of giving rise to oscillatory solutions resembling solar
cycles.  However, when we try to study the irregularities of the
cycles, the two models give completely different results.  We need
to introduce fluctuations to cause irregularities in the cycles.
In the high diffusivity model, fluctuations spread all over the
convection zone in about 5 years.  On the other hand, in the low
diffusivity model, fluctuations essentially remain frozen during
the cycle period.  Thus the behaviours of the two models are totally
different on introducing fluctuations.  
Over the last few years, several independent arguments have
been advanced in support of the
high diffusivity model (Chatterjee, Nandy \& Choudhuri 2004;
Chatterjee \& Choudhuri 2006; Jiang, Chatterjee \& Choudhuri 2007; Goel \&
Choudhuri 2009; Hotta \& Yokoyama 2010). We adopt the point of view
here the solar dynamo is most likely a high diffusivity flux transport
dynamo.

Three main sources of irregularities in dynamo models have been studied
by different authors over the years: (i) chaotic behaviours introduced
by nonlinearities of the dynamo process; (ii) fluctuations in the
generation of the poloidal field; (iii) fluctuations in the meridional
circulations.  The three following sections will focus on these three
sources of irregularities and discuss the question whether they can
cause grand minima.  Some of these sources of irregularities have been
investigated even before the flux transport dynamo model became popular,
by applying them to the earlier solar dynamo models.

\section{Effects of nonlinearities}

It is well known that nonlinear dynamical systems can show complicated
chaotic behaviours. Some of the earliest efforts of modelling solar cycle
irregularities invoked the idea of nonlinear chaos.  The full dynamo
problem is certainly a nonlinear problem in which the magnetic fields
produced by the fluid motions react back on the fluid motions.  The simplest
way of capturing the effect of this in a kinematic dynamo model (in
which the fluid equations are not solved) is to consider a quenching
of the $\alpha$ parameter as follows:
$$\alpha = \frac{\alpha_0}{1 + |\overline{B}/B_0|^2}, \eqno(2)$$
where $\overline{B}$ is the average of the magnetic field produced
by the dynamo and $B_0$ is the value of magnetic field beyond
which nonlinear effects become important.  There is a long history
of dynamo models studied with such quenching (Stix 1972; Ivanova \& Ruzmaikin 1977; Yoshimura 1978;
Brandenburg et al.\ 1989; Schmitt \& Sch\"ussler 1989). In most of the
nonlinear calculations, however, the dynamo eventually settles to a
periodic mode with a given amplitude rather than showing sustained
irregular behaviour.  The reason for this is intuitively obvious. 
Since a sudden increase in the amplitude of the magnetic field would
diminish the dynamo activity by reducing $\alpha$ given by (2) and thereby pull
down the amplitude again (a decrease in the amplitude would do the
opposite), the $\alpha$-quenching mechanism tends to lock the system
to a stable mode once the system relaxes to it. In fact, Krause \& Meinel (1988)
and Brandenburg et al.\ (1989) argued that the nonlinear stability
may determine the mode in which the dynamo is found.  Yoshimura (1978)
was able to reproduce some irregular features of the solar cycle by
introducing an unrealistic delay time of 29 years between the magnetic
field and its effect on the $\alpha$-coefficient.  In some highly
truncated models with the suppression of differential rotation, one
could find the evidence of chaos in limited parts of the parameter
space (Weiss, Cattaneo \& Jones 1984). K\"uker, Arlt \& R\"udiger (1999)
suggested that the quenching of differential rotation might have caused
the Maunder minimum.  This seems unlikely now on the ground that torsional
oscillations --- periodic modulations of differential rotation caused
by the dynamo-generated magnetic field --- appear like small perturbations.

It does not seem that the irregularities of solar cycles are primarily
caused by nonlinearities.  But that does not mean that nonlinearities
have no important consequences in the currently favoured flux transport
dynamo models.  In order to explain the even-odd or the Gnevyshev--Ohl
effect of solar cycles, Charbonneau, St-Jean \& Zacharias (2005) and
Charbonneau, Beaubien \& St-Jean (2007) made the highly
provocative suggestion that the solar dynamo
may be sitting in a region of period doubling just beyond the point of
nonlinear bifurcation. Recently the effects of nonlinearities introduced
by the quenching of turbulent diffusion (Guerrero, Dikpati \&  
de Gouveia Dal Pino 2009) and 
meridional circulation (Karak \& Choudhuri 2012) are being investigated.

\section{Fluctuations in poloidal field generation}

Since the mean field dynamo equations are derived by averaging over turbulence,
we expect fluctuations to be present around the mean. Choudhuri (1992) was
the first to suggest that these fluctuations will be particularly important
in the poloidal field generation. It is now difficult to believe that this
was an unorthodox and radical idea in 1992 when it was proposed, though 
this idea was explored further by Moss et al.\ (1992),
Hoyng (1993) and Ossendrijver, Hoyng \& Schmitt (1996).  
This idea was applied to the flux transport dynamo by
Charbonneau \& Dikpati (2000).

Let us consider the question how fluctuations in poloidal field generation
arise in the flux transport dynamo.  The Babcock--Leighton mechanism of poloidal
field generation depends the tilts of bipolar sunspot pairs.  While the average tilts
are given by Joy's law, one finds a large scatter around this average.  Longcope \&
Choudhuri (2002) provided a theoretical model of this scatter on the basis of the
idea that the rising flux tubes are buffeted by turbulence in the \cz.  This scatter
around Joy's law produces fluctuations in the poloidal field generation process
and we identify this as a primary source of irregularities in the solar cycle.
It may be noted that Choudhuri, Chatterjee \& Jiang (2007) and 
Jiang, Chatterjee \& Choudhuri (2007) modelled the last few cycles
by assuming the fluctuations in poloidal field generation to be the main
source of irregularities in solar cycles and predicted that the forthcoming
cycle~24 will be weak.  This prediction was based on the high diffusivity 
model.  There are enough indications by now that the upcoming cycle is going
to be a weak one, providing further support to the high diffusivity model.
 
\begin{figure}
\center
\includegraphics[width=14cm]{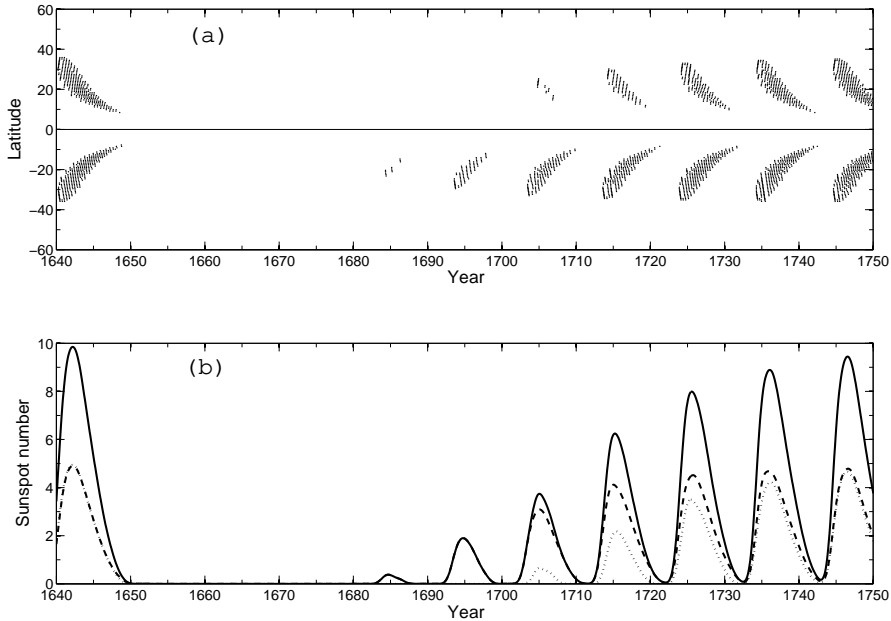}
\caption{The theoretical model of the Maunder minimum from Choudhuri \&
Karak (2009). This is based on a simulation in which the poloidal field
was reduced to 0.0 and 0.4 of its average value in the two hemispheres.
The upper panel shows the butterfly diagram.  The dotted and dashed lines
in the lower panel are sunspot numbers in the northern and southern hemispheres,
whereas the solid line is their sum.}
\end{figure}

We now come to question whether fluctuations in the poloidal field generation
can produce grand minima.  Several authors found that intermittencies resembling
grand minima can be obtained in simple dynamo models by introducing fluctuations
(Schmitt, Sch\"ussler \& Ferriz-Mas 1996; Mininni, Gomez \& Mindlin 2001;
Brandenburg \& Spiegel 2008). The effect of such fluctuations on flux transport
dynamo models has been investigated only recently. 
Charbonneau, Blais-Laurier \& St-Jean (2004) carried out a simulation by introducing
100\% fluctuations in $\alpha$ (the poloidal field generation parameter)
in a flux transport dynamo with low diffusivity.  They found intermittencies in
their simulations resembling grand minima. The low diffusivity of their model
ensured that the diffusive decay time or the `memory' of the dynamo was rather
long (of the order of a century) and most probably this long
memory played a role in producing intermittencies of similar duration. The
important question is whether the high diffusivity dynamo model, which we consider
to be the appropriate model for explaining solar cycles and which has a much
smaller diffusive decay time, can also produce similar intermittencies on introducing
fluctuations in poloidal field generation.  This question has been studied
by Choudhuri \& Karak (2009).

Choudhuri, Chatterjee \& Jiang (2007) proposed a simplified procedure for incorporating
the cumulative effect of fluctuations in poloidal field generation.  Since these
fluctuations in poloidal field generation would make the poloidal field at the end
of a cycle different from the average poloidal field one would get from the mean field
equations without including fluctuations, they suggested that the poloidal field
at the end of the cycle may be modified suitably to account for the cumulative effect
of the fluctuations. Choudhuri \& Karak (2009) found that the dynamo is pushed
into a grand minimum if the poloidal field at the end of a cycle falls to 0.2 of
its average value.  At the time of this work during the depth of a long sunspot
minimum, there was considerable speculation whether the Sun was entering another
grand minimum.  Choudhuri \& Karak (2009) concluded that the poloidal field had
fallen to only about 0.6 of its average value and hence the Sun should {\em not} be
entering another grand minimum.  On making the poloidal field in the northern and
southern hemispheres fall to respectively 0.0 and 0.4 of its average value, Choudhuri
\& Karak (2009) found that many characteristics of the Maunder minimum were
reproduced.  Fig.~1 shows the theoretical plots of butterfly
diagram and sunspot number, which compare favourably with the 
corresponding observational plots given
in Fig.~1(a) of Sokoloff \& Nesme-Ribes (1994) and Fig.~1 of Usoskin,
Mursula \& Kovaltsov (2000). We find that the theoretical model reproduced 
the fact that the Maunder minimum started abruptly, but ended gradually. It is
basically the growth time of the dynamo which determines the duration of the grand
minimum during which the magnetic field has to grow up again to return to normalcy.
As pointed out in \S~3, the operation of the dynamo during the grand minimum
presumably depends on the $\alpha$-effect and the theoretical model shows an ongoing
but subdued cycle of magnetic field in the solar wind. We sum up the theoretical
results in the following words. If the poloidal field at the end of a cycle turns out
to be very weak due to fluctuations in its generation process, then that can push
the dynamo into a grand minimum, from which it recovers gradually in the dynamo
growth time.  

\section{Fluctuations in meridional circulation}

It is well known that the period of the flux transport dynamo varies roughly as
the inverse of the \mc\ speed.  The period of the dynamo is approximately given
by the time taken by \mc\ at the bottom of the convection zone to move from
higher latitudes to lower latitudes.  In other words, the period of a flux
transport dynamo does not depend too much on the details of poloidal field generation
mechanism.  Probably this is the reason why the period of the dynamo during
a grand minimum like the Maunder minimum does not change drastically, even
though the poloidal field generation mechanism may be different from normal times
as explained \S~3. 

Since the \mc\ determines the period of the flux transport dynamo, 
it is obvious that any fluctuations
in \mc\ would have an effect on the flux transport dynamo. It has been found
recently that the \mc\ has a periodic variation with the solar cycle, becoming
weaker at the time of sunspot maximum (Hathaway \& Rightmire 2010; Basu \& Antia 2010).
Presumably the Lorentz force of the dynamo-generated magnetic field slows down
the \mc\ at the time of the sunspot maximum.  Karak \& Choudhuri (2012) found
that this quenching of \mc\ by the Lorentz force does not produce irregularities
in the cycle, provided the diffusivity is high as we believe.  We disagree with
the model of Nandy, Mu\~noz-Jaramillo \& Martens (2011) which assumes that the 
\mc\ changes abruptly at each sunspot maximum.  Our point of view is that the
periodic variation of \mc\ due to the Lorentz force cannot be responsible for
solar cycle irregularities and we need to consider other kinds of fluctuations
in \mc.

We have reliable observational data on the variation of \mc\ only for a little
more than a decade.  To draw any conclusions about the variation of \mc\ at
earlier times, we have to rely on indirect arguments. If we assume the cycle
period to go inversely as \mc, then we can use periods of different past solar
cycles to infer how \mc\ has varied with time in the last few centuries. On the
basis of such considerations, it appears that the \mc\ had random fluctuations
in the last few centuries with correlation time of the order of 30--40 years (Karak \& Choudhuri 2011).
We now come to question what effect these random fluctuations of \mc\ may have
on the dynamo.  Based on the analysis of Yeates, Nandy \& Mckay (2008), we can
easily see that dynamos with high and low diffusivity will be affected very
differently.  Suppose the \mc\ has suddenly fallen to a low value. This will
increase the period of the dynamo and lead to two opposing effects.  On the one
hand, the differential rotation will have more time to generate the toroidal field and
will try to make the cycles stronger.  On the other hand, diffusion will also have
more time to act on the magnetic fields and will try to make the cycles weaker.
Which of these two competing effects wins over will depend on the value of diffusivity.
If the diffusivity is high, then the action of diffusivity is more important and
the cycles become weaker when the \mc\ is slower.  The opposite happens if the
diffusivity is low.

As we pointed out in \S4 and \S6, there are enough indications that the diffusivity
of the solar dynamo is high.  If that is the case, then a slowing
of the \mc\ would make the cycles weaker.  Karak (2010) found that the flux transport
dynamo can be pushed into a grand minimum if the \mc\ drops to 0.4  of its normal
value.  This is another possible mechanism for producing a grand minimum. 

\begin{figure}
\center
\includegraphics[width=9cm]{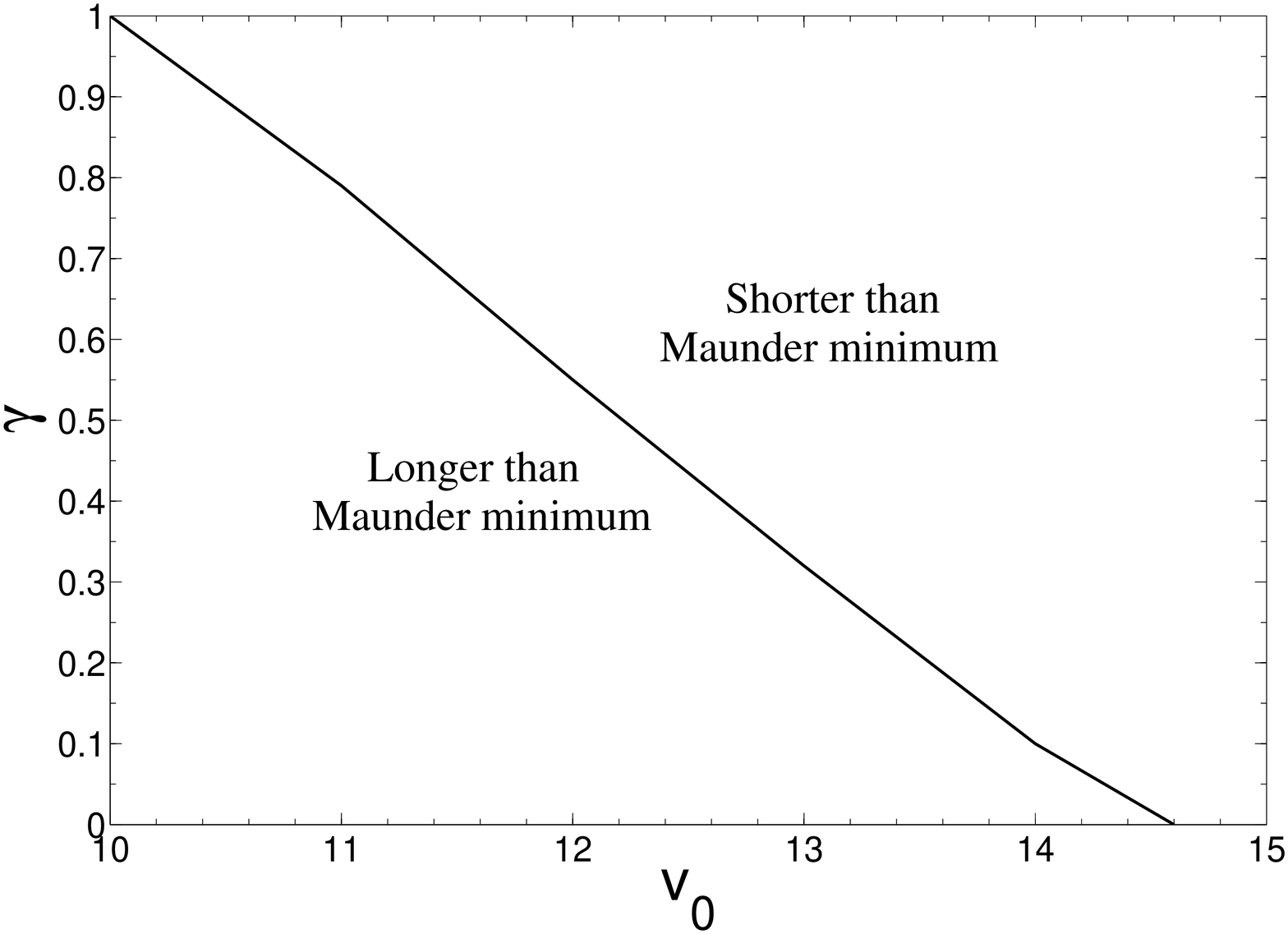}
\caption{The parameter space indicating the reduction factor $\gamma$ of the
poloidal field and the amplitude of the \mc\ needed to produce Maunder-like
grand minima.}
\end{figure}

Miyahara et al.\ (2004) found that cycles during the Maunder minimum became
somewhat longer, indicating that the \mc\ must have slowed down.  This supports
the theoretical idea that the weakening of \mc\ might have played an important
role in producing the Maunder minimum.  It should be noted that the opposite
would happen in the low diffusivity model.  The weakening of \mc\ and the lengthening
of cycles in a low diffusivity model should be associated with a grand maximum,
since longer cycles would allow the differential rotation to generate stronger
fields in the low diffusivity model.

\section{Concluding remarks}

Although there are many uncertainties in our theoretical understanding of grand
minima, it appears that fluctuations in poloidal field generation and fluctuations
in \mc\ are the main causes of irregularities in solar cycles and can also produce
grand minima.  We believe the solar dynamo to be a high diffusivity dynamo in
which a fall in the \mc\ makes cycles weaker. If fluctuations in poloidal field 
generation alone and fluctuations in \mc\ alone are to produce grand minima, then
the poloidal field at the end of a cycle or the \mc\ has to fall to rather low
values to cause grand minima.  The situation is a little less constrained if we
consider simultaneous fluctuations in both. Fig.~2 taken from Karak (2010)
shows the values to which
the poloidal field at the end of a cycle and the \mc\ have to fall if a grand
minimum is to be caused by their simultaneously falling to low values. This seems
to be the most likely scenario we have at the present time for explaining grand
minima. One important question is whether we can estimate how often this is likely
to happen.  Can we explain why there were 27 grand minima in the last 11,000 years?
We are looking at this question right now.

I end by summarizing what appears to me to be the most plausible theoretical
scenario for grand minima based on the flux transport dynamo.
Due to fluctuations in poloidal field generation and \mc, if both of them
simultaneously happen to become sufficiently weak, that may push the Sun into a grand minimum.
Within the grand minimum, the dynamo keeps operating on the basis of the $\alpha$-effect
and ultimately bounces out of the grand minimum in the dynamo growth time.
 
\section*{Acknowledgments}

My participation in IAU Symposium 286 was made possible by a JC Bose Fellowship 
awarded by Department of Science and Technology, Government of India.

\end{document}